\documentclass[12pt]{article}
\usepackage[
backend=biber,
style=ieee,
citestyle=numeric,
]{biblatex}
\usepackage[linguistics]{forest}
\usepackage{adjustbox}
\usepackage{latexsym}		
\usepackage{graphicx}
\usepackage{times}
\usepackage{url}
\usepackage{makecell}
\usepackage{array}
\usepackage{float}
\title{A bibLaTeX example}
\title{A Review of Attacks Against Language-Based Package Managers }

\author{Aarnav M. Bos \\
	CODE University of Applied Sciences \\
	Berlin, Germany \\
	{\tt aarnav.bos@code.berlin} \\}

\date{}

\begin{document}
	\maketitle
	\begin{abstract}
	The liberalization of software licensing has led to unprecedented re-use of software. Alongside drastically increasing productivity and arguably quality of derivative works, it has also introduced multiple attack vectors. The management of software intended for re-use is typically conducted by a package manager, whose role involves installing and updating packages and enabling reproducible environments. Package managers implement various measures to enforce the integrity and accurate resolution of packages to prevent supply chain attacks. This review explores supply chain attacks on package managers. The attacks are categorized based on the nature of their impact and their position in the package installation process. To conclude, further areas of research are presented.\\ 
		\\
		\\
		\\
		\\
		\\
		\\
		\\
		\\
		\\
	\end{abstract}
	\section{Introduction}
	Projects that reuse software may benefit from abstractions to common, complex problems, functionality that is written with established standards in mind, and faster time to market \cite{sommerville_software_2011}. As projects evolve and dependencies grow, the manual management of dependencies across mutiple systems becomes increasingly complicated. Package managers facilitate reproducible environments, the installation, updating and removal of software artefacts \cite{abate2011mpm}.
	\par
	Package managers have become essential to software engineering, so much so that languages such as Python \cite{developers_pip_nodate}, Go \cite{go_mod}, Rust \cite{cargo_pkg_mgr} and Javascript \cite{npm_a_pkg_mgr} provide them as a core piece of their language suite. 
	With developers opting to use even the most trivial packages\cite{abdalkareem2020impact}, the significance of package managers makes them attractive targets for attackers. 
	\\
	A software supply chain encompasses all parties and processes involved in constructing and delivering a final software product \cite{nygard_sok_2022}. This includes but is not limited to, package managers, package repositories, package developers and maintainers of package repositories. 
	The recent demonstrations of high impact attacks on package managers such as pip, Bundler and Yarn \cite{birsan_dependency_2021, johnson_picking_nodate}, which allowed attackers to discreetly install incorrect packages, ushered package managers into the spotlight, highlighting their importance in the supply chain.
	
	This review's contribution is an overview of supply chain attacks on package managers. It demonstrates three different families of attacks:
	Package Injection, where an attacker's goal is to install unintended packages, Denial of Service, where an an attacker's goal is to cause disruption in service and Code Injection, where an attacker's goal is to execute arbitrary code. Build scripts are excluded as an attack vector due to the nature of build systems allowing network access, file system access and code execution. This review also assumes that package repositories are secure and not operated by malicious parties.

	\section{Overview of Package Installation}
	The package installation process is separated into five different parts to categorize attacks distinctly. To install or update a package, a package manager queries the repository for metadata and attempts to find a matching version considering project constraints. If a matching version is found, it fetches, extracts and installs the artefacts. If configured, the package manager appends package metadata to its lockfile, a file tracking a list of packages and their metadata to enable reproducible environments.

	\subsection{Resolution}
	When a user requests their package manager to install a package, the package manager queries remote repositories to fetch meta-information about the package. Packages are queried by their name, which is their unique identifier on repositories. Considering constraints such as language version and existing package installations, the manager attempts to calculate a compatible version of the package; factoring in its dependencies and sub-dependencies. A suitable version match is not guaranteed. If a user is installing using a lockfile, the resolution step is not necessary as the lockfile explicitly provides the resolution data. When updating or removing packages, resolution is also required to make sure no other packages are affected in case of shared dependencies.
	
	\subsection{Fetching}
	Depending on the results of resolution, package artefacts are fetched from a remote repository or the local cache. The artefacts are typically zipped to save bandwidth. The package manager unzips them and stores them. Some package managers store packages in a directory that is accessible by all applications implemented in the target programming language while others store it in a per-project basis. Some can do both.
	
	\subsection{Verification}
	The verification of an installed package's integrity and authenticity is crucial. The verification of package integrity is typically done through a checksum, fetched alongside package metadata. A hash function is run on the package to see if the resulting checksum matches the one provided. In case of a Man In The Middle(MITM) attack, a false checksum can be provided, making integrity checking insufficient to assert validity. One way to assert authenticity is done through the use of public-key cryptography, where a package maintainer signs the provided checksum which the package manager subsequently verifies.

	\subsection{Installation}
	After fetching package artefacts, depending on the language, and package, an installation step is required. Packaging formats either implicitly or explicitly provide instructions on how to build and install a package in the form of a build script. Alongside build scripts, some package managers support pre and post installation scripts which are used to check for operating system dependencies, clean up build artefacts or for fetching additional resources, amongst other things.
	
	\subsection{Acknowledgement}
	If a successful installation took place, a package manager acknowledges it by informing the user. Some package managers, if configured, keep track of the list of dependencies and their resolution information in the form of a lockfile. The lockfile is automatically updated upon installation, updating or removal of dependencies.

	\section{Attack Types}
	\begin{table}[H]
		\label{tab:my-table}
		\begin{tabular}{|c|c|c|}
			\hline
			\textbf{Attack Name}                      & \textbf{Installation Step} & \textbf{Family}                  \\ \hline
			Man In The Middle                & Fetching          & Package Manipulation     \\ \hline
			Misconfigured Integrity Check    & Verification      & Package Manipulation     \\ \hline
			Misconfigured Authenticity Check & Verification      & Package Manipulation     \\ \hline
			Dependency Confusion             & Resolution        & Package Manipulation     \\ \hline
			Lockfile Tampering               & Resolution        & Package Manipulation     \\ \hline
			Zip-Bombs                        & Fetching          & Denial Of Service        \\ \hline
			Predictable Build Artifacts      & Installation      & Denial Of Service        \\ \hline
			Command Injection                & All               &  Code Injection \\ \hline

		\end{tabular}
			\caption{Taxonomy of Attacks}
	\end{table}
	The attacks are classified based on the position in the package installation process and the outcome of exploitation.

	\subsection{Package Manipulation}
	A vulnerability in a package manager which an attacker can utilize to install an unintended package is called Package Manipulation. Package manipulation can occur when the authenticity and integrity of artefacts are not verified or if the resolution of packages is done in an unpredictable manner. 
	\par
	With the fetching of package artefacts from remote repositories, there are at least three ways a package manager can be vulnerable to an MITM attack \cite{conti_survey_2016}:
	\begin{itemize}
		\item If the transmission does not use SSL/TLS to secure communications, it can be vulnerable to a IP-spoofing based MITM attack.
		\item If the validity of the SSL/TLS certificate used is not verified, it can be vulnerable to a SSL/TLS MITM Attack.
		\item If the validity of the response from the DNS server used to resolve the repository is not verified, it may be vulnerable to a DNS-spoofing based MITM attack.
	\end{itemize}
	 MITM attacks can enable an attacker to provide false or malicious package artefacts that are assumed to be legitimate by the package manager.
	\par
	Correct verification of the authenticity and integrity of the package artefacts is crucial. Invalid or absent verification of package artefacts' authenticity and integrity can allow an attacker to provide malicious artefacts, through an MITM attack for example, which may be accepted as valid ones \cite{athalye2014package, cve2013_1629}.
	\par
	Disagreements between parsers of data interchange formats such as JSON \cite{seriot_parsing_nodate} and XML \cite{spath_sok_nodate} are of significant concern to package management \cite{johnson_picking_nodate, tal_why_nodate} as many package managers use lockfiles to have predictable package resolution and reproducible environments. Lockfiles are updated automatically and manual changes are discouraged. Research by Johnson and Appelt \cite{johnson_picking_nodate}, and Tal \cite{tal_why_nodate} highlight how seemingly harmless manual changes to lockfiles can lead to package injection due to package managers parsing their lockfiles in an ambiguous or invalid manner. Yarn, when provided with duplicated attributes in a package entry, takes the last attribute \cite{noauthor_gitlabcom_nodate}. For example:
\begin{figure}[H]
	\centering
\begin{verbatim}
   corepack@^0.14.1:
        version "0.14.1"
        resolved "https://registry.com/"
        # duplicated field
        resolved "https://malicious.registry.com/" 
        integrity sha512-xyz
        # duplicated field
        integrity sha512-xyz-malicious 
\end{verbatim}
\end{figure}
This allows an attacker to change the package source and integrity, allowing them to install any package instead of the intended.
	Depending on the lockfile format, modifications conducted when adding, updating or removing a package can be large in terms of lines. Without extensive auditing, an application developer may assume the changes to the lockfile as automatic and proceed to integrate them, thus compromising their application. 
	
	Package managers must also be extremely cautious when updating the structure of package metadata, especially relating to dependencies as it has been known to have unintended consequences. Cargo, for example, introduced functionality which allowed projects to override dependency names by defining an alias for the dependency in the project declaration file \cite{cargo_}. The feature would allow projects to use the alias in place of the original name in their code; however, Cargo versions prior to the introduction of the feature parsed but ignored the alias. This could allow attackers to find projects which utilize aliases and create packages with those aliases on the package repository, leading anyone with an older version of Cargo to fetch unintended packages \cite{cargo_1_26_wrong_dep}. 
	\par
	Many package managers have a default public repository they query for packages when a user wants to install or update a package. They also allow users to specify custom repositories as organizations and users may wish to have private packages. Dependency Confusion is a vulnerability where package names from private repositories are duplicated on public repositories, causing confusion in resolution for the package manager \cite{birsan_dependency_2021}. Pip and Bundler, for example, resolved to whichever package had the highest version amongst it's sources \cite{cve2018_20225, cve2020_36327}, allowing an attacker to trick the package manager into fetching their package over the intended one.
	\subsection{Denial Of Service}
	A denial of service attack is an attack where the perpetrator attempts to cause disruption in a service by attempting to disable acess to or exhaust a victim's computational resources \cite{hussain_framework_2003}. This may include network bandwidth, memory, storage or computing power. Package managers which do not limit the amount of data they extract from compressed package artefacts can be vulnerable to a zip bomb \cite{zip_bomb_cargo}. A zip bomb is a malicious archive file which can crash a computer by overflowing its memory or disk space, or by putting excessive load on it's CPU \cite{zipbomb}. 
	\par
	Package managers may create temporary directories or files if necessary during the package build process \cite{exttracting_cargo, cve2013_1888, gocache}.
	If the names of these artefacts are predictable and if the package manager is unable to overwrite the file or directory, it may lead to denial of service \cite{cve2014_8991}.
	Furthermore, if package managers follow symbolic links when creating predictable build artefacts, a malicious actor could create symbolic links to sensitive files or directories\cite{cve2013_1888, exttracting_cargo}; allowing them to corrupt arbitrary files and potentially causing a denial of service attack. As such, a cryptographically secure random name must be assigned to all temporary artefacts to prevent such an attack. 

	\subsection{Code Injection}

	Command injection is a subset of a Code Injection \cite{inj} attack where the perpetrator attempts to execute arbitrary commands on a victim's computer by abusing how an application executes shell commands \cite{commandinjection}. It results from unsanitized input \cite{commandinjection}. Some package managers support a git \cite{git} repository as a package source. Since git is a command-line application, package managers use shell commands to automate git processes. An attacker may provide a malicious URL for a git repository which can lead to a command injection when utilized by the package manager \cite{cve_2022_36069, ruby_cmd_inect}. As package managers deal with foreign input such as package names, metadata and artefacts across all processes, rigorous sanitization must be implemented to prevent command injection.

	\section{Conclusion}
	An overview of attacks on package managers, including proven demonstrations is presented and the attacks are categorized by their outcome. Despite the importance of package managers in the software supply chain, a lack of systematic research is evident. The next steps in research could be:
	\begin{itemize}
		\item Reproducible environments - How do package managers parse and handle lockfiles. Do they adhere to the interchange specification?
		\item A thorough overview of package verification processes implemented in package managers, their efficiency and potency.
		\item A modern taxonomy of package management
	\end{itemize}
	\medskip
	\printbibliography
\end{document}